\def \SAIT #1 #2 {{\em Mem.\ Soc.\ Astron.\ It.\/} {\bf #1}, #2}
\def \MESS #1 #2 {{\em The Messenger\/} {\bf #1}, #2}
\def \ASTRNACH #1 #2 {{\em Astron. Nach.\/} {\bf #1}, #2}
\def \AAP #1 #2 {{\em Astron. Astrophys.\/} {\bf #1}, #2}
\def \AAL #1 #2 {{\em Astron. Astrophys. Lett.\/} {\bf #1}, L#2}
\def \AAR #1 #2 {{\em Astron. Astrophys. Rev.\/} {\bf #1}, #2}
\def \AAS #1 #2 {{\em Astron. Astrophys. Suppl. Ser.\/} {\bf #1}, #2}
\def \AJ #1 #2 {{\em Astron. J.\/} {\bf #1}, #2}
\def \ANNREV #1 #2 {{\em Ann. Rev. Astron. Astrophys.\/} {\bf #1}, #2}
\def \APJ #1 #2 {{\em Astrophys. J.\/} {\bf #1}, #2}
\def \APJL #1 #2 {{\em Astrophys. J. Lett.\/} {\bf #1}, L#2}
\def \APJS #1 #2 {{\em Astrophys. J. Suppl.\/} {\bf #1}, #2}
\def \APSS #1 #2 {{\em Astrophys. Space Sci.\/} {\bf #1}, #2}
\def \ASR #1 #2 {{\em Adv. Space Res.\/} {\bf #1}, #2}
\def \BAIC #1 #2 {{\em Bull. Astron. Inst. Czechosl.\/} {\bf #1}, #2}
\def \JSQRT #1 #2 {{\em J. Quant. Spectrosc. Radiat. Transfer\/} {\bf #1}, #2}
\def \MN #1 #2 {{\em Mon. Not. R. Astr. Soc.\/} {\bf #1}, #2}
\def \MEM #1 #2 {{\em Mem. R. Astr. Soc.\/} {\bf #1}, #2}
\def \PLR #1 #2 {{\em Phys. Lett. Rev.\/} {\bf #1}, #2}
\def \PASJ #1 #2 {{\em Publ. Astron. Soc. Japan\/} {\bf #1}, #2}
\def \PASP #1 #2 {{\em Publ. Astr. Soc. Pacific\/} {\bf #1}, #2}
\def \NAT #1 #2 {{\em Nature\/} {\bf #1}, #2}
\newcommand{\gtrsim}{\mathrel{\hbox{\rlap{\hbox{\lower4pt\hbox{$\sim$}}}\hbox{$>$}}}
}
\newcommand{\ltrsim}{\mathrel{\hbox{\rlap{\hbox{\lower4pt\hbox{$\sim$}}}\hbox{$<$}}}
}

\documentstyle{memsait}
\input epsf.sty
\begin{opening}
\title{TRANSIENT X--RAY SOURCES  OBSERVED with the RXTE ALL-SKY MONITOR 
after 3.5 YEARS}
\author{HALE BRADT, ALAN M. LEVINE, RONALD A. REMILLARD, DONALD A. 
SMITH}
\institute{$^1$Massachusetts Institute of Technology, Room 37-587, Cambridge MA 02139-
4307 USA}
\date{} 
\end{opening}

\begin{document}

\oddpagefooter{}{}{} 
\evenpagefooter{}{}{} 
\ 
\bigskip

\begin{abstract}

We present light curves of a sample of ``transient'' sources observed with the All-Sky 
Monitor (ASM) of the Rossi X-ray Timing Explorer (RXTE). The light curves extend over 3.5 
years. They are presented in three groups:  six neutron-star systems, eight
black-hole-candidate systems, and an additional diverse set of six objects that are either 
transient sources (in the sense of usually being undetectable) or persistent sources showing
transient behavior. The outburst profiles of these sources show
reproducible characteristics within one source and from source to
source, as well as large variations. These profiles together with the
profiles of the hardness ratios from the ASM are a valuable
resource for the understanding of accretion instabilities. We summarize briefly some recent 
work by observers on these somewhat arbitrarily selected sources. 

\end{abstract}

\section{Overview}
For a period of 3.5  years (at the time of this workshop), the All-Sky Monitor (ASM; Levine 
et al. 1996) on RXTE has been monitoring the entire sky for new (uncataloged) transient x--ray sources while also recording the intensities of the known (cataloged) sources  The 
current catalog contains about 325 source positions of which about 180 have yielded positive 
detections. The monitoring of a given source has been reasonably continuous except for times 
when the sun is relatively close to the source and except for a period of $\sim$7 weeks 
shortly after launch when the detectors were turned off due to a temporary breakdown 
problem. The detected sources include many well known persistent sources as well as
a substantial number of `transient' sources. Some of these are
recurrent and others are in their first known outburst. Most of
the latter have been discovered in the RXTE era, either with other
satellites, e.g. CGRO and BSAX, or with RXTE. A few were discovered prior to the launch of 
RXTE. 

Some of the RXTE discoveries have been made during scans with the highly-sensitive, but
narrow-field-of-view, PCA instrument while the others have been made
with the ASM. In almost all of these cases, whether the transient is
new or recurrent or whether it was discovered with RXTE or not, the
PCA has carried out pointed observations, and the ASM data have been
analyzed to provide a light curve (including upper limits) of the
source that extends back to early 1996. The exceptions are faint sources ($\ltrsim$25 mCrab) 
discovered with the PCA during scans of the crowded galactic-center region. Source confusion 
renders some of these inaccessible to the ASM.

The number of transient sources thus far observed by RXTE (PCA and
ASM) now exceeds 40, excluding Be star systems. We consider a transient to be a source 
that, for significant periods, has been below the detection level of the proportional-counter 
experiments (a few millicrabs) and which shows at least an order of magnitude change of 
intensity that exceeds the one-day ASM threshold ($\sim$15 mCrab at $3\sigma$ for a 
cataloged source). A more precise definition will be required for quantitative studies of 
transient occurrence rates. Also, a complete search to these levels requires care because our 
thresholds vary with position on the sky and with time because of variable source density and 
variable sky exposure.  

Our most recent tabulation (November 1999) shows 23 `new' transients studied with RXTE 
during the first known outburst, or series of outbursts. Among these, sixteen were
discovered with RXTE (9 with the ASM and 7 with the PCA). In addition there are 19 
recovered transients, i.e., previously known sources recovered after a period of quiescence. 
The RXTE/ASM has played a substantial role in the recoveries of these transients. A 
recovery often leads to the discovery of important new characteristics of the source. An 
earlier (1997) tabulation of transient sources observed with the RXTE/ASM may be found in 
Remillard (1999). An updated version will be released later. 
 
The occurrences or reoccurences of transients and also the changes of
state of persistent sources provide valuable opportunities for the
study of the processes that modulate x-ray emission. The early
detection and study of these events require a relatively sensitive
all--sky monitor together with a highly sensitive narrow--field x--ray
instrument that can be pointed rapidly to any point in the sky. The
RXTE combines these capabilities for the first time in the history of
x--ray astronomy. Thus new progress on the understanding of the
accretion processes in these variable sources is now becoming
possible.

\section{ASM Light Curves}
Light curves of transient sources from RXTE/ASM data show examples of repetitive 
behavior as well as considerable variation within the same source,
e.g. from outburst to outburst. There are `failed outbursts'
interspersed with `normal' outbursts. There are outbursts that usually
turn completely `off' (less than a few mCrab), but sometimes remain luminous
at a low level for long periods. Outburst profiles include `fast rise with
slow exponential-like decay' sometimes interspersed with flat-topped
and more irregular profiles. These light curves provide a quantitative challenge to models of 
accretion disk instabilities and of the capture of stellar wind from the companion.

In Figures 1--6, we present sample light curves (1.5 -- 12 keV) for the 3.5--year
period of RXTE's operations. The intensities are given in ASM ct/s where 75 ct/s corresponds 
to the x-ray intensity of the Crab nebula. Most of the light curves are displayed in the form of 
1--day averages of the 10 to 30 intensity measures typically obtained per day. The dates are 
given in MJD where:

1996 Jan. 0.0 = MJD 50082.0

1997 Jan. 0.0 = MJD 50448.0

1998 Jan. 0.0 = MJD 50813.0

1999 Jan. 0.0 = MJD 51178.0.

Figure 1 shows the light curves for six neutron-star binary sources.
Figures 2, 3, and 4 show the light curves for eight
black-hole-candidates, while Figs. 5 and 6 show six additional
sources of diverse types.  Most of these sources are transients in the
classic sense that they are not detectable above some threshold for
long periods of time. Figure 5 includes sources that we would not label as transients, but 
which do exhibit transient behavior in some sense of the word. 

The light curves shown are for the entire bandwidth of the ASM
detectors, namely 1.5 -- 12 keV. In fact, the data are telemetered and
stored in three energy channels, 1.5 -- 3 keV, 3 -- 5 keV, and 5 -- 12
keV, which can serve as a further tool in the study of transient
sources. The curves are shown here on a rather compressed scale;
larger scales reveal substantially more detail. The intensities are available in both graphical 
and numerical form on the internet sites at MIT and GSFC, e.g.,  
http://heasarc.gsfc.nasa.gov/xte\_weather/ and http://space.mit.edu/XTE/ASM\_lc.html.

Here, we point out features in the light curves of selected transients that should be pertinent 
to the classification of transient types. The characteristics of
the previously known sources are well documented in van Paradijs
(1995). The discovery references for a number of the new or recovered
sources are listed in Remillard (1999). 

We also point the reader to papers that can serve as an introduction to the more recent 
literature for the sources we have selected for illustration. This serves to give the reader a 
flavor of the research carried out by RXTE in the past few years. Much of this research has 
been stimulated by results from the ASM, e.g., because it provides notice of a changing state. 
This presentation is intended to point out the potential of such data for the understanding of 
accretion processes.  

\section{Neutron-star binaries; Fig. 1}

4U 0115+63. This HMXB (high-mass x--ray binary) is a 3.6--s pulsar. Recently its optical 
counterpart has been reclassified as an O9e (Unger et al. 1999). In 1999 at MJD $\sim$51250 
(hereafter simply 51250), it underwent a major outburst. Studies with BSAX and RXTE 
during this event led to the discoveries of high-harmonic cyclotron lines (Heindl et al. 1999; 
Santangelo et al. 1999). The outburst profile in the ASM light curve is quite symmetric. In 
addition, there were several weak outbursts in 1996 (50300) whose peaks are spaced by 
multiples of the 24--d orbital period. In larger scale plots they appear to have profiles similar 
to the larger outbursts. 

X 1608--522. This well known LMXB (low-mass x--ray binary) is an x--ray
burster as are many LMXBs. The source was in outburst when RXTE was launched and 
erupted again at 50850. The profile of
the latter outburst shows a remarkably very fast rise, a rapid initial
descent, a hesitation at about half maximum followed by a less rapid
fall to the lower quiescent level. The source was undetectable ($\ltrsim$10 mCrab) for almost
a year (50500 -- 50800) before the latter outburst. In contrast, it exhibited a sustained, 
variable low--level flux at 0.03 -- 0.1 Crab for a year or more after each of the outbursts. This 
could be an indicator that the source is intrinsically unstable between on and off states. This 
low level (for the ASM) is several orders of magnitude greater than
the quiescent luminosity of $\sim10^{33}$ erg/s. See Rutledge et al. (1999) for a discussion 
of the quiescent state. 

3A 1942+274. This Ariel source was recovered for the first time in many
years by the ASM in late 1998. It was found to be a 16-s pulsar in
RXTE PCA observations immediately upon its recovery (Smith \&
Takeshima 1998). The previously suggested optical counterpart (Israel, Polcaro \& Covino 
1998) is now considered to be an unlikely candidate (Israel, pvt. comm.). The irregularly 
spaced multiple peaks in the light curve may represent enhanced accretion near periastron in 
an  $\sim$80-d elliptical orbit (Campana, Israel \& Stella
1999). The source remains active through at least mid November 1999.

XTE J2123--058. This previously unknown source at high galactic
latitude ($­-36\deg$) was discovered in ASM data (Levine, Swank \& Smith 1998). It is an 
atoll LMXB with twin x-ray kHz oscillations and x-ray bursts (Homan et al. 1999). An optical 
counterpart with V = 17 and a 6--h orbit was immediately discovered (Tomsick et al. 1999a) 
with orbital modulations that continued beyond the x-ray outburst (Soria, Wu \& Galloway 
1999). The outburst at 51000 is quite weak (85 mCrab); it exhibits the same hesitation during
the decay seen in 1608--522. Possible precursor activity $\sim$100 d before
the peak is apparent at $\sim$15 mCrab, but may be an artifact from x-ray Solar 
contamination.

Rapid Burster. The ASM data show 6 outbursts of activity since the RXTE launch. They 
reaffirm the $\sim$200-d previously-noted quasi-periodicity of these outbursts, but for the first 
time show the extent and evolution of each outburst. Each outburst lasts about 5 weeks and, 
in PCA data, exhibits two phases, the first of which is characterized by Type I 
(thermonuclear) bursts and the second by Type II (accretion) bursts (Guerriero et al. 1999). 

Aql X-1. This well-known atoll, bursting, LMXB recurring transient has recently had its 
optical identification (V1333 Aql) clarified. The counterpart is the western component of a 
0.5-arcsec double with V= 21.6 in quiescence and with a late K-star classification (Chevalier 
et al. 1999). ASM detections of outburst states made possible studies of kHz oscillations in 
the active state that include demonstrations of (1) a sudden decrease in frequency and flux 
after an x-ray burst (Yu, Zhang \& Zhang 1999) and (2) low-energy lags of $\sim$1 radian 
consistent with a Doppler-delay model (Ford 1999). 

The ASM light curve of Aql X-1 shows 5 outbursts, including those at the 
beginning and the end of the plot and another two ``failed'' bursts at 50270 
and 51200. The intervals between these events range from 200 to 300 d. 
Two of the outburst light curves (50700 and 50900) are roughly symmetric 
in shape with comparable rise and fall times. The other two have faster 
rises than decays. The failed outbursts may indicate that the conditions for 
outburst are marginal, as we suggest for the low-level persistent flux from 
x1608--522. Aql X-1 lies on the thermal-viscous instability boundary 
derived by van Paradijs (1996) for neutron stars. 

\section{Black-hole candidates; Figs 2, 3, 4}

4U1630--47. This long-known transient and black-hole candidate (from
soft-hard spectral components) exhibits outbursts roughly in accord
with the reported $\sim$600--700 d intervals (Kuulkers et al. 1997). Three 
outbursts are seen in the ASM data with separations $\sim$700 d and 
$\sim$450 d. The longer interval follows the outburst with the largest 
integrated flux. All three outbursts exhibit sharp maxima at the leading 
and trailing edges of the profile, and the leading edges rise faster than the 
trailing edges decay. In spite of this, the latter two outbursts have 
different shapes; they appear to have failed to attain the full profile of the 
first outburst. The flux remains markedly above threshold after the second 
outburst. Again this behavior could be an indicator of marginal instability. 

The 1996 outburst (50200) revealed deep narrow (minutes) x-ray 
absorption dips in RXTE/{PCA data (Kuulkers et al. 1998).  This outburst 
is included in the historical review of outburst behavior in 4U 1630--47 
(Kuulkers et al. 1997). This review points out the heretofore unrecognized 
complexity of the outburst behavior of this source  The evolution of the x-
ray spectral components of the 1998 outburst (50850) from BeppoSAX are 
presented by Oosterbroek et al. (1998).

GRO J1655--40. This is a well established black-hole system with a
dynamical mass for the collapsed object of $\sim$6--7 solar masses (Orosz \&
Bailyn 1997, Shabaz et al. 1999). It exhibits superluminal radio jets which establish it
as a ``microquasar'' (Tingay et al. 1995). First discovered in 1994,
it was solidly below threshold prior to and after the $\sim$450--d 1996--1997
outburst. The optical turn-on was found to precede the x-ray emission
by $\sim$5 days (Orosz et al. 1997). Observations with the PCA exhibited several QPOs 
including 300 Hz when the source spectrum was particularly hard (Remillard et
al. 1999b). The evolution of x-ray spectral components is given by 
Mendez, Belloni \& van der Klis (1998), Tomsick et al. (1999b), and 
Sobczak et al. (1999a). Echo mapping (x-ray to optical) locates the 
reprocessing region to be in the accretion disk rather than the mass donor 
star (Hynes et al. 1998). 

XTE J1748--288. This RXTE/ASM-discovered transient (Smith, Levine \& Wood 1998) 
exhibited a single outburst at 51000 with a very rapid rise ($\sim$2 d) and slow
exponential-like decay. It was above threshold for 60 d and was detected to at least 100 keV 
with BATSE. (Harmon et al. 1998).  PCA observations showed QPO at 0.5 and 32 Hz (Fox 
\& Lewin 1998). The spectral and QPO evolution has been studied by Revnivtsev, 
Trudolyubov \& Borozdin (1999). Transient radio emission (Hjellming et al., 1998, Fender \& 
Stappers 1998) became extended after 10 days. The jet motion at $>$20 mas/d indicated an 
apparent speed $>$ 0.93c for a $\geq$8 kpc distance. Later the speed decreased and the 
leading edge of the radio jet brightened due to a shock in the interstellar medium, the first 
such shock known in a galactic source.  There is no reported optical counterpart; its galactic 
coords. are 0.7, --0.2).

XTE J1755--324. This source, also ASM discovered (Remillard et al. 1997), was noteworthy 
for its extremely soft color in the ASM (HR2 = 0.3) which suggests black-hole candidacy. 
(HR2 is the harder color of the two ASM x-ray energy colors.) The intensity profile was 
similar to XTE J1748--288 in that it rose in $\sim$2 d and decreased
quasi-exponentially. However, unlike 1748--288, but reminiscent of the
neutron-star system 1608--522 (Fig. 1), it halted its descent after
about 50 days, increased slightly to a second maximum and then descended 
to non-detectability after a total time of $\sim$105d. The temporal and 
spectral evolution is described by Revnivtsev, Gilfanov \& Churazov (1998) 
who find it to be a ``canonical'' x-ray nova such as Nova Muscae 1991. 
Again there is no reported optical identification at this writing; galactic 
coords. 358.0, --3.6.

XTE J2012+381. This transient, discovered with the ASM (Remillard, 
Levine \& Wood 1998), had a hard initial spike and also reached a very 
low hardness ratio (HR2 = 0.4). In ASCA data it had an ultrasoft 
spectrum with a hard tail suggesting a black hole (White et al. 1998). A 
radio counterpart was detected (Hjellming \& Rupen 1998a, Pooley 1998). 
A V = 21.3 star (1.1 arcsec from a foreground 18th magnitude star) lies 
within 0.4 arcsec of the radio source and is tentatively identified as the 
optical counterpart (Hynes et al. 1999). The light curve is strongly double 
peaked with an additional low maximum 150 d after the onset.

GX 339--4. This long-known black-hole candidate with a persistent
x-ray flux entered a bright and soft (HR2 $\sim$0.3) state beginning at
$\sim$50800 and returned to its low state $\sim$400 d later. It is not generally considered a 
``transient''; nevertheless the infrequent high soft states could be considered so. The ASM 
high/soft state is accompanied by a low high-energy (BATSE) flux and a marked reduction of 
radio emission (Fender et al. 1999a). The transition is reminiscent of the 1996 Cyg X-1 
transition to its high/soft state (Wen et al. 1999 and refs. therein). The temporal and spectral 
characteristics during the rise and at maximum showed a ``typical'' high/soft state (Belloni et 
al. 1999b). See also the multiwavelength (radio, optical, x-ray) studies in a 3-paper series 
(e.g., Smith, Filippenko \& Leonard 1999). 

XTE J1915+105 (Fig. 3). This remarkable ``microquasar'' (Mirabel \&
Rodriguez 1994) has been active continuously since the launch of RXTE. The x--ray light 
curves show what is probably the most dramatic variability of any known x-ray
source. Extensive PCA observations (vertical-line markers at top) have
permitted the definition of a number of distinct accretion states (e.g., Greiner, Morgan \& 
Remillard 1997) This source also exhibits superluminal radio jets (e.g. Fender et al. 1999b). 
A dynamical mass does not exist for  this source, but its high luminosity strongly implies a 
black-hole compact object. The object is replete with unusual x-ray oscillatory modes that 
repeat on time scales of tens to thousands of seconds (e.g., see Muno et al. 1999). One such 
mode has been associated with micro-radio outbursts with high confidence (Pooley \& Fender 
1997, Eikenberry et al. 1998, Mirabel et al. 1998), thus linking specific accretion 
configurations with jet creation. These events appear to be associated with the loss via 
accretion of the inner portion of the accretion disk (see Belloni et al. 1997, Pooley \& Fender 
1997). The radio--x-ray association is also seen on longer time scales in Fig. 3 where 
increased radio fluxes are sometimes associated with low hard x--ray states, e.g. at 50750, 
with, in addition, bright flaring at the onset and end of such low states. The source is replete 
with low frequency x--ray QPOs at frequencies that vary with time and also
a 67-Hz QPO that reappears when the spectrum is
hard (Morgan, Remillard \& Greiner 1997). Recent work has addressed the interplay of the 
QPOs and the variation of the spectral components, e.g. Muno et al. (1999) and Markwardt, 
Swank \& Taam (1999).

XTEJ1550--564 (Fig. 4). This was the first previously-unknown very--bright transient found 
in the RXTE era. It was discovered in ASM data with new triggering software soon after its 
onset (Smith 1998). This permitted PCA observations during the rise, the times of which are 
marked in the figure. The x-ray light curve with its narrow peak, flat plateau and second 
maximum is not typical of a prototype x-ray nova, e.g., 0620--00. It reaches 6.8 Crab 
brightness in the dramatic brief flare at $\sim$51050. X-ray emission was observed to 200 
keV with BATSE (Wilson et al. 1998). The evolution of the x-ray spectral components in the 
PCA and ASM data were tracked by Sobczak et al. (1999b). The source was found in the 
very high, high/soft, and intermediate canonical outburst states of black-hole x-ray novae. X-
ray QPOs were abundant from 0.05 to 185 Hz (Cui et al. 1999; Remillard et al. 1999a). An 
optical counterpart brightened about 4 magnitudes over the quiescent B $\approx$22 state 
(Jain et al. 1999). The extinction E(B--V) = 0.7 indicated a distance of 2.5 kpc and a 
luminosity consistent with a K0--K5 star (Sanchez-Fernandez, et al. 1999). A likely radio 
counterpart was found (Campbell-Wilson et al. 1998). 

\section{Various sources; Fig. 5}
Here we illustrate the diversity of x-ray variation in a few other selected
sources. Most would not be deemed transients under the conventional
definition. However they all exhibit transient behavior in the broader sense.

CI Cam. Here we see the remarkably brief outburst from the symbiotic
star CI Cam, discovered with the ASM (Smith et al. 1998). The rise took place over a few 
hours, and the exponential-like fall (Fig. 6) showed a time scale that began at $\sim$0.5 d 
and later became $\sim$2.3 day (Belloni et al. 1999a). It was
above the ASM threshold for only 9 days. The ASM detection enabled studies with ASCA, 
BSAX and the RXTE PCA as well as in the radio and optical. This object was reported to 
have corkscrew radio jets reminiscent of SS433 (Hjellming \& Mioduszewski 1998). Belloni 
et al. (1999a) suggest this is a Be-star + neutron-star binary system but do not exclude a 
black-hole; see Belloni et al. for references to several studies of this outburst. 

GS 2138+568 (=Cep X-4?). This source was discovered in 1972 with OSO--7. It was 
rediscovered with Ginga which found it to be a 66-s pulsar. It is believed to be a Be-star 
system with a total of four recorded outbursts. A study with BATSE and RXTE (Wilson, 
Finger \& Scott 1998) of outbursts in 1993 and 1997 also serves as a review of the literature. 
In the ASM data, the source remained above threshold for about 40 days. It is among the 
fainter sources ($\sim$30 mCrab) detectable with the ASM in 1--day averages.

4U 1705--44. This modestly bright LMXB burster with no known optical
counterpart exhibits large intensity variations that have rise and
fall times that quite consistently are on the order of 50 days.  This
source has historical `off' states but has essentially always been detectable since Feb. 1996 
in the ASM data. It also exhibits radio emission. Kilohertz QPO have been discovered with 
RXTE (Ford, van der Klis \& Kaaret 1998). The fast timing behavior at lower frequencies 
based on the EXOSAT archive are reported by Berger \& van der Klis (1998). 

SMC X-1. This well-known high-mass system with an eclipsing (3.9 d) x--ray pulsar (0.71s)
reveals itself conclusively to have a quasi 60-d period in the ASM data. It is probably a 
precessing disk system (Wojdowski et al. 1998).  The pulsar continues to show a steady 
spin-up rate (Kahabka \& Li 1999).

Cyg X-3. This unusual binary, with a 4.8-h x--ray period (probably orbital) exhibits 
spectacular radio flares. Its ASM x--ray light curve exhibits one sustained flaring interval and 
several shorter ones. Bright radio flares occur during the sustained active period, with peak 
intensities of 3 to 9 J at wavelength 13 cm at 50465, 50485, and 50610 (Ogley et al. 1998). 
The orbital period evolution has recently been revisited by Matz (1997).

Mkn 501. This blazar exhibits a year-long period of x-ray activity in the ASM data from 50400 
to 50800. This active period was accompanied by dramatic flux variations and a high average 
flux ($\sim$1.4 Crab) at TeV energies (Quinn et al. 1999; Aharonian et al. 1999). It was on 
MJD 50607 that an intense rapid (hour scale) TeV flare was detected (Quinn et al.). 
Aharonian et al. report an intensity correlation with ASM data with zero time lag during this 
active period. In 1998 the TeV flux had decreased to $\sim$20\% of the Crab. The ASM data 
serve to guide TeV observers in target selection. The light curve (4-day averages) shows the 
capability of the ASM for the study of the brighter AGN sources which, even so, are quite 
weak in the ASM. Mkn 501 reaches 35 mCrab in this plot.

\section{Conclusion}
The light curves shown here are illustrate the richness of the ASM data and improve the 
prospect for better understanding of the accretion processes underlying transient behavior. 
The outburst profiles, the marginally-on states, the durations of on and off states, and the 
hardness parameters should all serve as diagnostics of these processes. A planned 
reanalysis of the entire data base with improved procedures should provide a uniform search 
for transients sources down to $\sim$7 mCrab at positions away from the galactic center. 
With good fortune the RXTE/ASM will continue to provide comprehensive light curves to the 
community for at least several more years.

\acknowledgements
The authors are grateful to the RXTE/ASM team at MIT and GSFC and to NASA for support 
under Contract NAS5--30612

\begin{figure}
\epsfysize=14cm
\hspace{3.5cm}\epsfbox{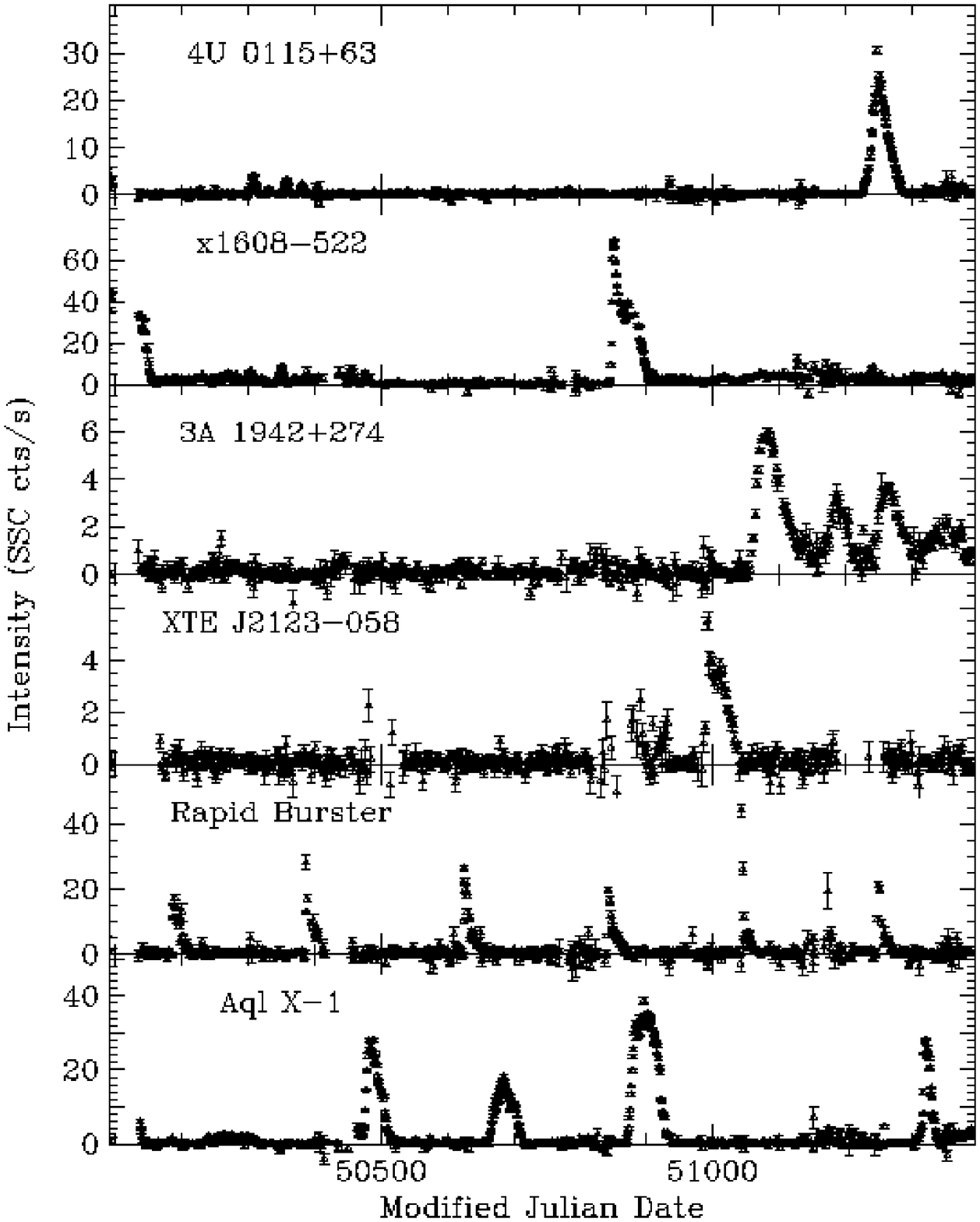}
\vspace{2cm}
\caption[h]{RXTE/ASM light curves for six neutron-star binary-system
transients. The data points represent 1-day averages of the 10 -- 20
(typical) daily measurements in the 1.5--12 keV band. 75 ct/s
corresponds to 1.0 Crab. MJD 50082 = 1996 Jan. 0.0.}
\end{figure}

\begin{figure}
\epsfxsize=11cm
\hspace{3.5cm}\epsfbox{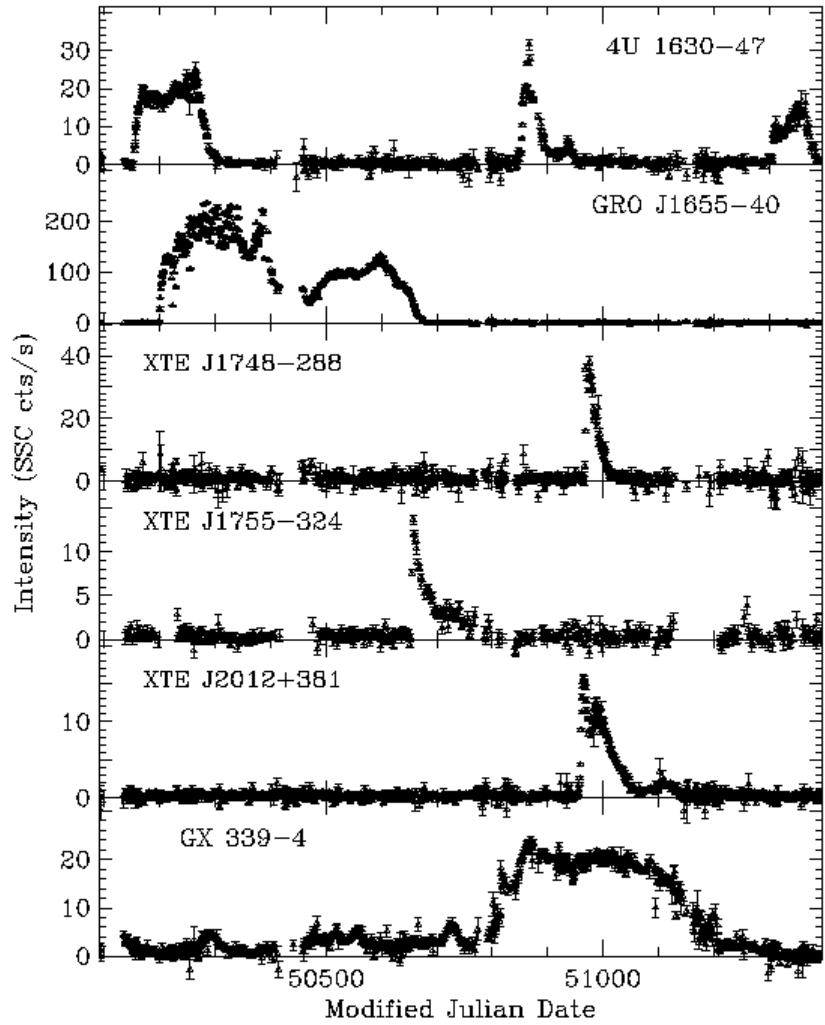}
\vspace{2cm}
\caption[h]{RXTE/ASM light curves for six black-hole binary-system
transients. GX 339--4 shows a transient bright soft state; it also
exhibits ``off'' states. See also caption to Fig. 1}
\end{figure}

\begin{figure}
\epsfxsize=15cm
\hspace{3.5cm}\epsfbox{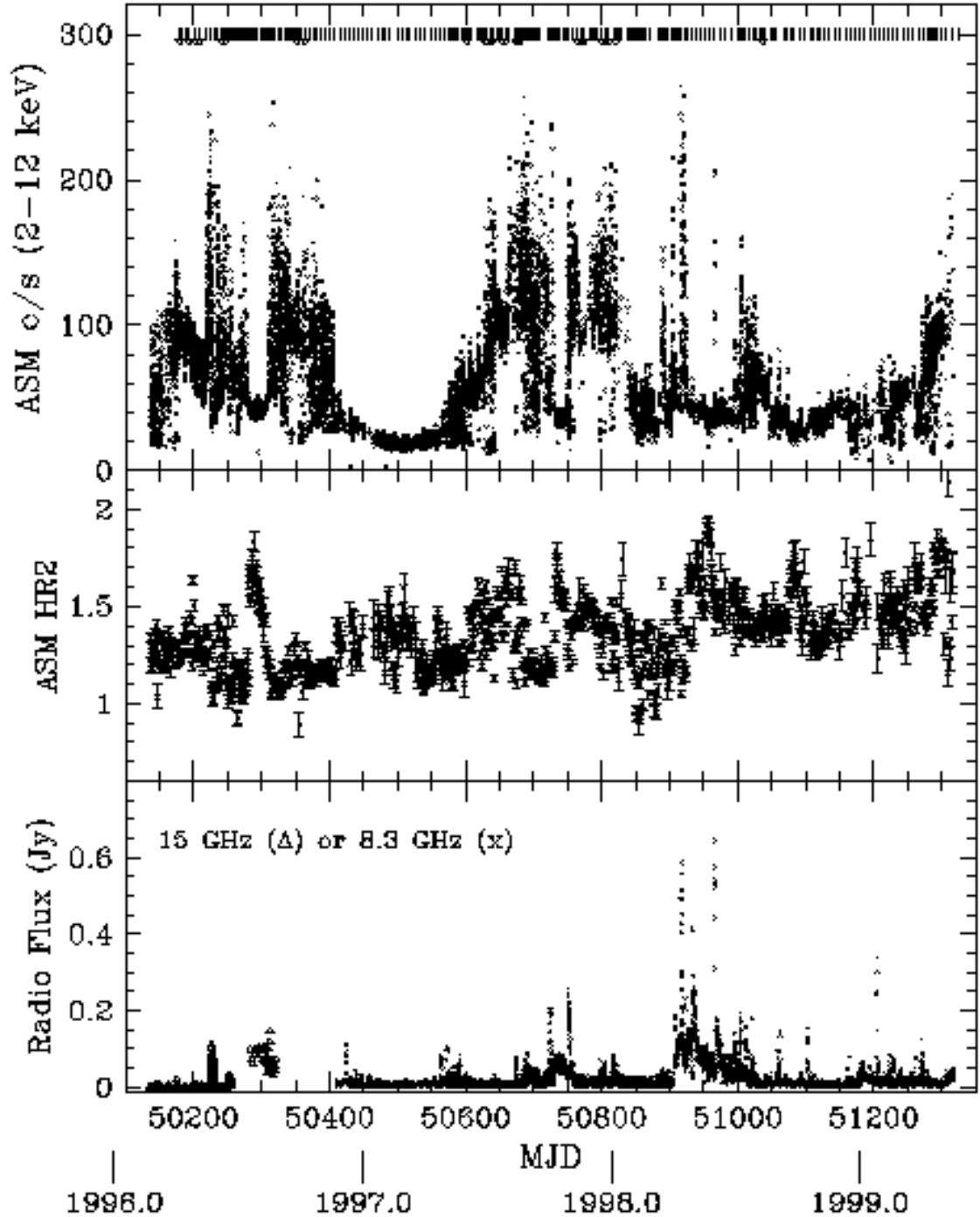}
\vspace{1cm}
\caption[h]{RXTE/ASM x-ray light curve, hardness ratio, and radio
fluxes for the ``microquasar'' GRS 1915+105. The data points for the
x-ray light curve are individual 90-s dwells which show the huge
variability on short time scales. The hardness ratios (5 --12 keV)/(3 -- 5
keV) are one-day averages. The markers at the top indicate times of
PCA observations and the less frequent occurrence of the 67-Hz QPO.}
\end{figure}

\begin{figure}
\epsfxsize=15cm
\hspace{3.5cm}\epsfbox{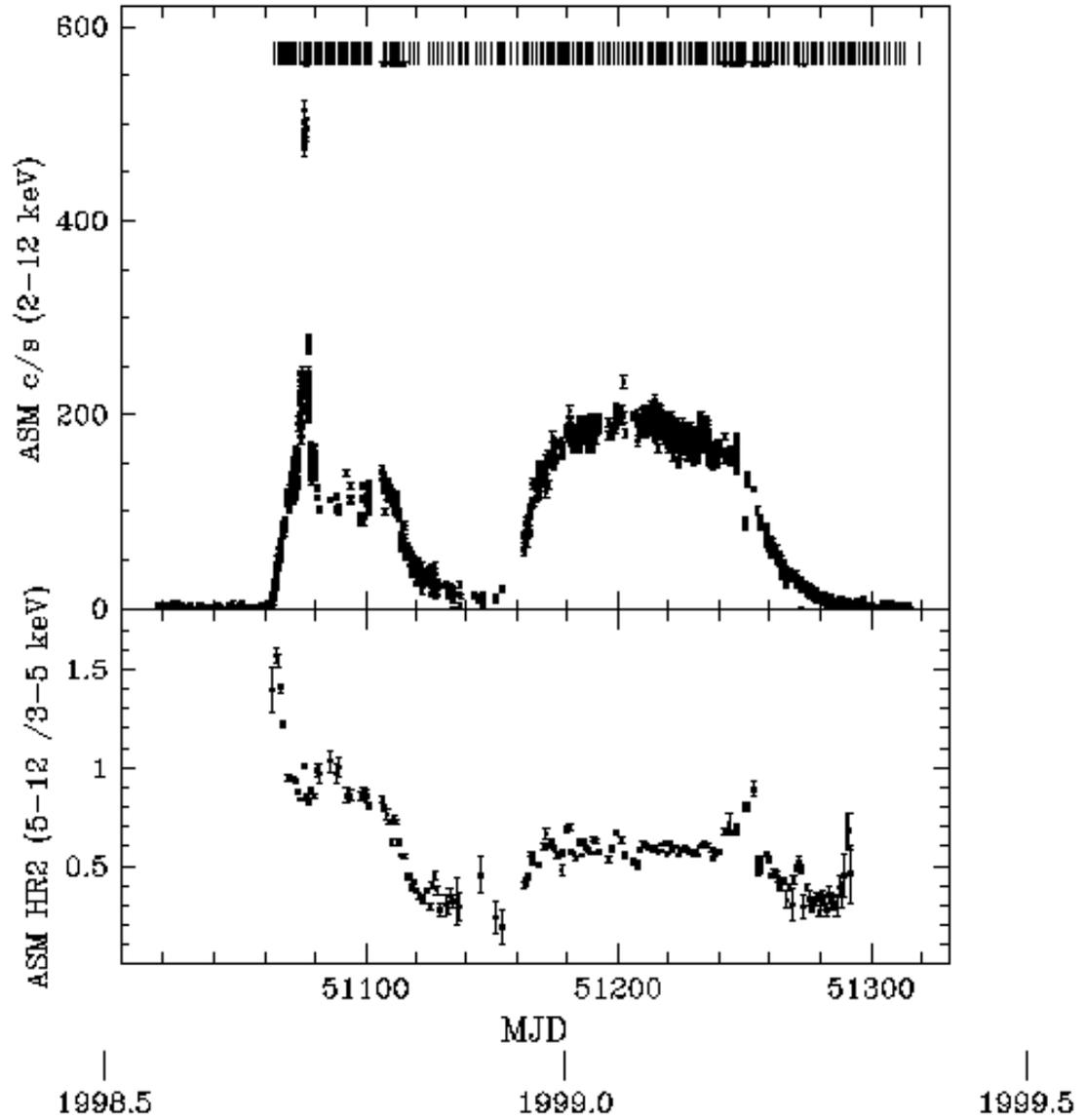}
\vspace{2cm}
\caption[h]{RXTE/ASM light curve and hardness ratio for XTE J1550--564
which reached an intensity of 6.8 Crabs. The markers indicate the PCA
observation times and the less frequent detections of the 185 Hz
QPO. The intensity points are 90-s dwell objects and the hardness
ratios are one-day averages.}
\end{figure}

\begin{figure}
\epsfxsize=11cm
\hspace{3.5cm}\epsfbox{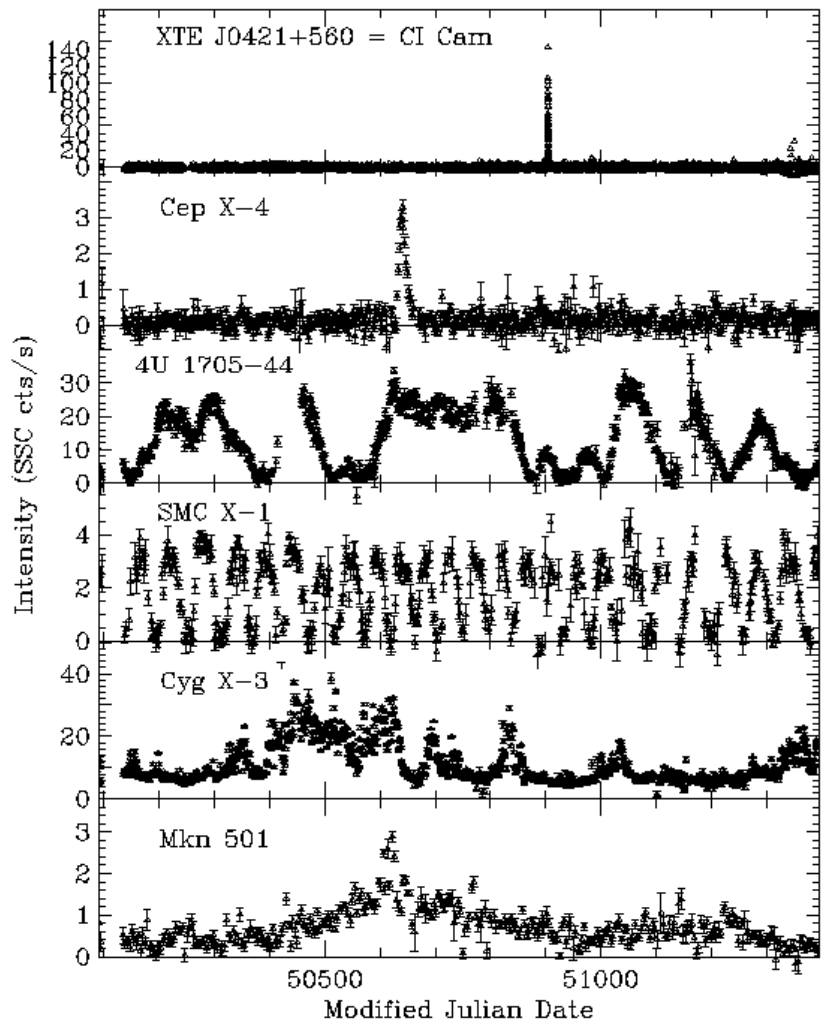}
\vspace{3cm}
\caption[h]{RXTE/ASM light curves for six sources of diverse character
and diverse transient behavior. The CI Cam points are from individual
90-s dwells, those for Mk 501 are 4-d averages, and the others are 1-day averages. See also
caption to Fig. 1}
\end{figure}

\begin{figure}
\epsfxsize=13cm
\hspace{3.5cm}\epsfbox{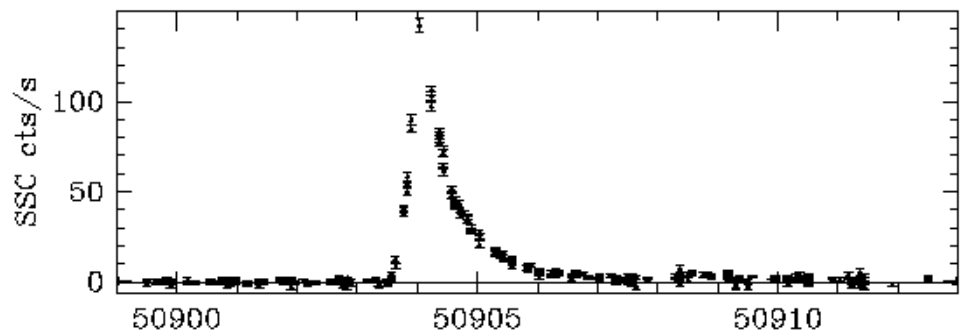} 
\caption[h]{RXTE/ASM expanded light curve for CI~Cam illustrating its rise in hours and $\sim$1-d descent.}
\end{figure}

\end{document}